\documentclass[preprint,showkeys,aps,amssymb,prd]{revtex4}  
\usepackage{bm,graphicx} 
\usepackage{epsf} 
\newcommand\beq{\begin{equation}} 
\newcommand\beqa{\begin{eqnarray}} 
\newcommand\beqan{\begin{eqnarray*}} 
\newcommand\eeq{\end{equation}} 
\newcommand\eeqa{\end{eqnarray}} 
\newcommand\eeqan{\end{eqnarray*}} 
\newcommand\bcrit{b_{\rm c}} 

\newcommand\gravr{{\sf m}_\bullet}

\newcommand\order[2]{\mathcal{O}\left[{#1}^{#2}\right]} 
\newcommand\rr{\bar{r}} 
\newcommand{\alphahat}{\hat{\alpha}}

\begin{document} 
\title{Light's Bending Angle due to Black Holes:  From the Photon Sphere to Infinity} 
\author{S.  V.  Iyer} 
\affiliation{Department of Physics \& Astronomy, SUNY Geneseo,\\ 
1 College Circle,\\ Geneseo, NY 14454;\\   
{\tt iyer@geneseo.edu}} 
\author{A.  O. Petters} 
\affiliation{Departments of Mathematics and Physics,   Duke University,\\   
Science Drive,\\ Durham, NC 27708-0320;\\   
{\tt petters@math.duke.edu}\\ 
{\rm (Revised March 15, 2007) } } 

\begin{abstract} The bending angle of light is a central 
quantity in the theory of gravitational lensing. We develop an 
analytical perturbation framework for calculating the bending 
angle of light rays lensed  by a Schwarzschild black hole. 
Using a perturbation parameter given in terms of the gravitational 
radius of the black hole and the light ray's impact parameter, 
we determine an invariant series for the strong-deflection bending 
angle that extends beyond the  standard logarithmic deflection term 
used in the literature.   In the process, we discovered  an improvement
to the standard logarithmic deflection term.
Our perturbation 
framework is also used to derive as a consistency check, the 
recently found weak deflection bending angle series.  We also reformulate 
the latter series in terms of a more natural invariant perturbation 
parameter, one that smoothly transitions between the weak and strong 
deflection series.  We then compare our invariant strong deflection 
bending-angle series  with the numerically integrated exact formal 
bending angle expression, and find less than $1 \%$ discrepancy for 
light rays as far out  as  twice the critical impact parameter. The 
paper concludes by showing that the strong and weak deflection bending 
angle series together provide an approximation that is within $1 \%$
of the exact bending angle value for light rays traversing anywhere 
between the photon sphere and infinity. 
\end{abstract} 
\keywords{gravitational lensing, black holes} 
\maketitle 
\section{Introduction} 
\label{sec:intro} 

One of the early striking predictions 
of General Relativity is that the weak-deflection bending 
angle of star light grazing the Sun is of the form 
\beq 
\label{eq-GRweak-1} {\hat{\alpha}}_{\rm Einstein} (r_0)\, 
= \, 4\, \left(\frac{\gravr}{r_0}\right) \ 
+ \ \order{\left(\frac{\gravr}{r_0}\right)}{2}, 
\hspace{1in} \mbox{\small \sf (First-Order Weak-Deflection)} 
\eeq 
which at leading order is twice the value given by 
Newtonian gravity. Eddington's 1919 confirmation of 
the leading term in (\ref{eq-GRweak-1}) was the first 
observation of gravitational lensing and brought Einstein's 
new gravitational theory instant scientific acclaim. Gravitational 
lensing in the weak-deflection limit has since been studied 
extensively, yielding numerous applications in astrophysics 
and cosmology (e.g., Schneider et al. 1992 
\cite{sef},  Petters et al . 2001 \cite{plw}, 
Kochanek et al. 2005 \cite{ksw}).  In addition, over 
the past two years  weak-deflection lensing has been 
employed to create tests accessible to current or 
near-future instruments,  of gravity theories such as 
PPN models  and $5$-dimensional, string-theory inspired, 
braneworld gravity (Keeton and Petters 2005-2006 \cite{kp1,kp2,kp3}).  
In \cite{kp1}, the first-order weak deflection formula (\ref{eq-GRweak-1}) 
was also extended to all orders in $\gravr/r_0$ and re-expressed 
as an invariant perturbation series (since $r_0$ is a coordinate 
dependent quantity \cite{coord-dep}). Interested readers may also 
find a recent analysis in \cite{DeLuca} of the weak deflection limit.

In recent years, however, the exciting promise of planned 
space-borne black hole imaging instruments has ignited  
research activity in the  analytical study of lensing in 
the strong-deflection regime (e.g, Virbhadra and 
Ellis 2000 \cite{ve}, Frittelli, Kling, and Newman 2000 \cite{fkn}, 
Eiroa, Romero, and Torres 2003 \cite{ert}, Petters 2003 \cite{ptt}, 
Perlick 2004 \cite{prk}, Bozza, Capozziello, De Luca, Iovane, 
Mancini, Scarpetta, and Sereno 2001-2005 \cite{bozzaetal}).   

For a Schwarzschild black hole of physical mass $M$, the 
spacetime geometry in the vicinity of the photon sphere 
at radius $r = 3 \gravr$, where $\gravr = G M/c^2$ is the 
black hole's gravitational radius, is revealed through the 
resulting strong-deflection gravitational lensing. In 1959, 
Darwin \cite{darwin}  computed the first-order term of the 
bending angle of light traversing deep inside the black hole's 
potential --- i.e., close to the photon sphere: 
\beq 
\label{eq-GRstrong-1} 
{\hat{\alpha}}_{\rm Darwin} (r_0)\, = \, - \pi \ 
+ \ 2 \, \log \left[\frac{36 (2 - \sqrt{3})\, \gravr}{r_0 \, 
- \, 3 \gravr}\right] \ + \ \order{h'}{}, 
\hspace{0.2in} 
\mbox{\small \sf (First-Order Strong-Deflection)} 
\eeq 
where $h' = 1 - (3 \gravr)/r_0$. He also showed analytically 
that near the photon sphere there are two families of relativistic 
images, which are images determined by light rays that loop around 
the photon sphere  at least once before reaching the observer. 
Other authors have confirmed this strong-deflection multi-looping 
lensing effect (e.g., Atkinson 1965 \cite{atkinson}, 
Luminet 1979 \cite{luminet},  Chandrasekhar 1982 \cite{chandra}, 
Ohanian 1987 \cite{ohanian}, several recent authors \cite{ve}-\cite{bozzaetal}). 
These studies were based on evaluating the lowest-order term (out from 
the photon sphere) of the light ray's strong-deflection bending angle. 
Equation (\ref{eq-GRstrong-1}) is the well-known leading 
logarithmic deflection term. 

In this paper, we develop a perturbative framework that allows us 
to generalize Darwin's strong deflection result (\ref{eq-GRstrong-1}) 
to any order in $h'$.  Surprisingly,  we also found that the leading
logarithmic 
deflection term employed in the literature can be improved. Earlier 
studies (e.g., \cite[p. 188]{darwin},  \cite[p. 132]{chandra}) arrived 
at the leading logarithmic expression by a perturbation scheme that seems
to combine higher and lower order terms. (We leave a definite assessment to
the judgment of the reader --- see the Appendix.)
By re-doing the perturbation theory and 
being careful to compare only terms of the same order, we obtain an 
improvement (i.e., more accurate expression) 
to the leading logarithmic deflection term. Furthermore, since $r_0$ is 
coordinate dependent \cite{coord-dep}, we re-formulate our strong-deflection 
bending angle in terms of a coordinate-independent series. In particular, we 
compute this invariant series explicitly to $3$rd-order in the perturbation 
parameter $b' = 1 - b_c/b$, where $b_c = 3\sqrt{3} \gravr$ is the critical 
impact parameter. Our perturbation framework was also used to compute the 
weak-deflection bending-angle series directly and we found it to be in complete 
agreement with the expansion found recently in \cite{kp1}.   Finally, we 
show that our invariant bending-angle series is  in excellent agreement 
with the numerically computed exact formal expression for the bending 
angle. This is done for both the strong and weak deflection  limits, 
and the span from the photon sphere to infinity. These results should be 
applicable to analytical lensing studies across these regimes and serve 
as a limiting case to check bending angle results in spacetime geometries 
generalizing the Schwarzschild metric. 

The outline of the paper is as follows: Section \ref{sec:exactschw} expresses 
the light's bending angle in a formal exact expression involving a difference 
of elliptic integrals of the first kind.   In Section~\ref{sec:photosphere}, we 
expand the  strong-deflection bending angle in terms of an invariant series 
going outward from the the photon sphere.  This section includes the 
improvement  
to the logarithmic term. Finally, Section~\ref{sec:comparison} gives a numerical 
comparison between the perturbative and exact bending angles across the range 
from the photon sphere to infinity. 

\section{Formal Exact Strong-Deflection Bending Angle} 
\label{sec:exactschw} 

A Schwarzschild black hole is the unique static, spherically symmetric, 
asymptotically flat vacuum solution of the Einstein equation.  The metric is 
given in Schwarzschild coordinates $(t,r,\theta,\phi)$ by 
\beq 
\label{eq-Sch- metric1}   
{\rm d}s^2 = - \left( 1 - \frac{2\gravr}{\rr}     \right) {\rm d}t^2     \ 
+ \ \left( 1 + \frac{2\gravr}{\rr}     \right)^{-1} {\rm d}\rr^2 + \rr^2\,     
(d\theta^2 + \sin^2 \theta \, d \phi^2), 
\eeq 
where $t = c \tau$ and $\gravr=G M/c^2$ (gravitational radius) 
with $\tau$ physical time and $M$ the physical mass of the black hole at the origin.  

Consider a standard gravitational lensing situation where a point source and 
observer  lie in the asymptotically flat region. 
In a typical lensing scenario, the source and observer are on opposite sides
of the black hole.  However, this restriction can be lifted  
in strong-deflection
lensing.  Suppose that the source is close to the optical axis 
passing through the observer and black hole. By spherical 
symmetry, it suffices to choose the source-to-observer light 
rays as lying in the equatorial plane ($\theta = \pi/2$). The Euler-Lagrange 
equations yield that the light rays are governed by (e.g., \cite{kp1}): 
\beq 
\label{eq-dphi-dr} 
\left(\frac{d \phi}{d r}\right)^2 = \frac{1}{ r^4 \, \sqrt{1/b^2 \ 
- \ ( 1 \ - \ 2\gravr/r )/r^2}}, 
\eeq 
where $b = |L/E|$ is the impact parameter with $L$ and $E$ the 
respective angular momentum and energy invariants of the light 
ray. Setting $u = 1/r$, re-write (\ref{eq-dphi-dr}) as 
\beqa 
\label{eq-cubic} 
\left(\frac{du}{d\phi}\right)^2 &=&u^4 \left[\frac{E^2}{u^4 L^4}
-\frac{1}{u^2} \left(1-2\gravr u \right)\right] \nonumber \\ 
&=& 2 \gravr u^3 -u^2 + {1\over{b^2}}.
\eeqa 
This cubic polynomial has a maximum of two positive roots and 
at most one negative root. 

Writing (\ref{eq-cubic}) as $$ B(u)=2\gravr (u-u_1)(u-u_2)(u-u_3), $$ we 
consider the case of one negative root $u_1$ and two distinct positive 
roots $u_2$ and $u_3$.  The three roots, given in terms of an intermediate 
constant $Q$ that allows us to line up the roots in the 
order $u_1<u_2<u_3$ are given by (e.g., p. 130 \cite{chandra}): 
$$ u_1 = \frac{r_0 -2\gravr-Q}{4\gravr r_0}, \quad  u_2 
= \frac{1}{r_0}, \quad u_3 = \frac{r_0 -2\gravr+Q}{4\gravr r_0}. $$ 
Here $r_0$ is the light ray's distance of closest approach, which is 
determined from (e.g., Eq. (12) of \cite{kp1}): 
\beq 
\label{r0tob} b^2=\frac{r_0^3}{r_0-2 \gravr}. 
\eeq 

By comparing the coefficients in $B(u)$ to those in the original 
polynomial in equation(\ref{eq-cubic}), we obtain the following two 
relations between $Q$ and the quantities $b, \gravr, r_0$: 
$$ \frac{Q^2-{(r_0 - 2 \gravr)}^2}{8 \gravr r_0^3} = \frac{1}{b^2}, $$ 
which is equivalent to $$ Q^2=(r_0 - 2 \gravr) (r_0+ 6 \gravr). $$ 

The bending angle of the lensed light ray is given by (e.g., Eq. (20) of \cite{kp1}): 
\beqan 
\alphahat &=& 2 \int_0^{1/r_0} \frac{du}{\sqrt{2\gravr(u-u_1)(u-u_2)(u-u_3)}}
-\pi\\ && \nonumber \\ 
&=& \sqrt{\frac{2}{\gravr}}\int_0^{1/r_0} \frac{du}{\sqrt{(u-u_1)(u_2-u)(u_3-u)}}-\pi 
\eeqan 
Split the above integral into two parts to make the lower limit equal to 
the smallest root $u_1$: 
\beq 
\label{eq-alphahat1} 
\alphahat =\sqrt{\frac{2}{\gravr}} \left[ \int_{u_1}^{u_2}
\frac{du}{\sqrt{(u-u_1)(u_2-u)(u_3-u)}} \ 
- \ \int_{u_1}^0 \frac{du}{\sqrt{(u-u_1)(u_2-u)(u_3-u)}} \right] \ - \ \pi. 
\eeq 

Now, the integrals in (\ref{eq-alphahat1}) can be realized as elliptic 
integrals of the first kind (see Byrd and Friedman \cite{byrd} for an 
introduction to elliptic integrals): 
$$ \alphahat =\sqrt{\frac{2}{\gravr}} \left[ \, 
\frac{2\, F (\Psi_1, k)}{\sqrt{u_3 - u_1}}  \ 
- \ \frac{2\, F (\Psi_2, k)}{\sqrt{u_3 - u_1}} \right] \ - \ \pi,$$ 
where $F(\Psi_i,k)$ 
is an incomplete elliptic integral of the first kind with amplitudes 
$$ \Psi_1 = \frac{\pi}{2},\quad \Psi_2 = \sin^{-1} \sqrt{\frac{-u_1}{u_2 - u_1}}, $$ 
and modulus $$ k^2= \frac{u_2 - u_1}{u_3 - u_1}. $$ Explicitly, 
$$ \Psi_2 = \sin^{-1} \sqrt{\frac{Q  + 2 \gravr - r_0}{Q  
+ 6 \gravr - r_0}}, \quad k^2 = \frac{Q - r_0 + 6\gravr}{2 Q}. $$ 
Hence, the exact bending angle simplifies to 
\beq 
\label{eq-exact-bangle} 
\alphahat=4\sqrt{\frac{r_0}{Q}} \left[K(k) \ 
- \ F(\Psi,k)\right]\ -\ \pi, 
\eeq 
where $K(k)$ and $F(\Psi,k)$ are the complete and incomplete 
elliptic integrals of the first kind, respectively, and $\Psi = \Psi_2$.  
As a check of (\ref{eq-exact-bangle}), note that in the limit 
when $\gravr \rightarrow 0$, we obtain $Q=r_0$, $k=0$ and $\Psi=\pi/4$, 
which in turn imply $K(k)=\pi/2$ and $F(\Psi,k)=\pi/4$ to give us 
zero deflection as expected. 

It is important to add that the exact bending angle 
expression (\ref{eq-exact-bangle}) serves mainly as 
a {\it formal} expression.  The latter has to be evaluated 
to obtain explicit analytical and physical properties about the 
nature of the bending angle. Equations (\ref{eq-GRweak-1}) due to 
Einstein and (\ref{eq-GRstrong-1}) due to Darwin are the first-order 
evaluations of (\ref{eq-exact-bangle}) in the weak and strong 
deflection limits, respectively. The challenge of course is in 
evaluating (\ref{eq-exact-bangle}) beyond those terms.  The 
weak-deflection series in terms of the impact parameter out
to many orders beyond (\ref{eq-GRweak-1}) was found 
recently in \cite{kp1}.   We shall now determine the strong-deflection 
series beyond (\ref{eq-GRstrong-1}), re-derive the weak series result 
in \cite{kp1}, and reformulate the weak series in terms of a new perturbation 
parameter to allow a seamless comparison covering the span from the 
strong to weak deflection limits --- i.e., from the photon sphere to infinity. 

\section{Expansion of Bending Angle Beyond the Photon Sphere} 
\label{sec:photosphere} 

The {\it photon sphere} is defined by the radius $r = 3 \gravr$, which 
marks an unstable photon orbit. Light rays that cross within the photon 
sphere are captured by the black hole (e.g., \cite{darwin,chandra}). Using 
the relations given in Section~\ref{sec:exactschw}, we can see that exactly 
on the photon sphere the impact parameter invariant is given by the 
critical value $b_c=3\sqrt{3}\gravr$. Photon orbits of interest 
to us are between $r_0=3 \gravr$ and $r_0=\infty$, and indeed 
our focus will be on the region closer to the photon sphere.  

We first express $h'$ in terms of $b'$. Equation (\ref{r0tob}) 
is a cubic in $r_0$ that is readily solved to yield: 
\beq 
\label{r0solveb} r_0 = \frac{2\, b}{\sqrt{3}} \, 
\cos\left[\frac{1}{3}\, \cos^{-1} \left(\frac{-3\sqrt{3} \, 
\gravr}{b}\right) \right]. 
\eeq 
\smallskip
The quantities $r_0$ and $b$ of course have very different 
physical meaning for the light rays.  Relative to an inertial 
observer at infinity, the quantity $r_0$ is the distance of 
closest approach to the center of the black hole, while the 
impact parameter $b$ is the perpendicular distance from the 
black hole's center  to the asymptotic tangent line to the 
light ray converging at the observer. Overall,  the 
quantity $r_0$ approaches $b$ as we extend into regions 
well beyond the photon sphere, but near the photon 
sphere, the values of $r_0$ and $b$ are different.

To seamlessly traverse from regions near the photon to 
those at infinity, a natural choice of invariant parameter 
is $$b'=1 \ -\ \frac{b_c}{b},$$ which ranges from $0$ at the 
photon sphere to $1$ at infinity (asymptotically flat region). 

\subsection{Affine Perturbative Form for Bending Angle} 

\label{subsec:affine} 

Our goal is to show that the strong-deflection bending angle 
can be expressed as an ``affine perturbation'' series in $b'$. More 
precisely,  define an {\it affine perturbation} series about a 
function $g$ as 
\beq 
\label{eq-affine-pert} 
f(x)=(A_0 +\cdots +A_p x^p+ \cdots)\, g(x)\ + \ (B_0 + \cdots + B_q x^q + \cdots), 
\eeq 
where $A_i$ and $B_i$ are constants with $p$ and $q$ positive 
rational numbers. We shall demonstrate in Section~\ref{subsec-inv-3rd} that 
the bending angle has an invariant affine perturbation series of the form 
\beqa 
\label{eq-affine-pert-angle} 
\alphahat(b') & = & \Bigl(\sigma_0 +  \sigma_1 \, (b') + \sigma_2 \, (b')^2 
+ \sigma_3 \, (b')^3 + \cdots \, \Bigr) \, 
\log{\left(\frac{\lambda_0}{b'}\right)}\nonumber \\ 
& & \hspace{0.5in}  \ + \ \Bigl(\rho_0  + \rho_1 \, (b') 
+ \rho_2 \, (b')^2 + \rho_3 \, (b')^3 + \cdots \, \Bigr), 
\eeqa 
where $\lambda_0$,  $\sigma_i$ and $\rho_i$ are numerical 
constants. Note that (\ref{eq-affine-pert-angle}) is not a 
Taylor series expansion because of the appearance of the 
logarithmic term.  However, we shall see that this logarithmic 
term is not exactly (\ref{eq-GRstrong-1}). 

\subsection{Bending Angle Series Beyond the Logarithmic Term} 
\label{subsec:Bangle-series-hp} 

We now consider the exact bending angle in the region around the photon 
sphere by expanding out from the photon sphere using $h=\gravr/r_0$. The 
expression for the bending angle can now be rewritten as a 
function of $h$ using the following useful relations: 
\beqan 
Q &=& r_0 \sqrt{(1-2h)(1+6h)}\\ \nonumber \\ 
k^2 &=& \frac{\sqrt{(1-2h)(1+6h)}-(1-6h)}{2\sqrt{(1-2h)(1+6h)}}\\ \nonumber \\ 
\Psi &=& \sin^{-1} \sqrt{\frac{\sqrt{(1-2h)(1+6h)}-(1-2h)}{ \sqrt{(1-2h)(1+6h)}-(1-6h)}}. 
\eeqan 
As $r_0$ increases from $3\gravr$ to $\infty$, the parameter $h$ goes 
from $1/3$ to $0$, which corresponds to the invariant impact parameter 
increasing from $3\,\sqrt{3} \gravr$ to $\infty$.  Intuitively, when the 
bending angle is computed in the regime near $b = 3\,\sqrt{3} \gravr$, we 
speak of {\it strong-deflection} (since we shall 
show in Section~\ref{subsec:Bangle-series-hp} that the bending 
angle can become arbitrarily large), while {\it weak-deflection} will 
refer to regions with large $b$.   

Before proceeding with the substitution of the above quantities in the 
expression for bending angle, we present an example of the standard 
series expansions for a complete elliptic integral of the 
first kind (e.g., see page 298 in \cite{byrd})
in terms of the modulus $k$:

\beq
K(k)=\frac{\pi}{2} \left[1+\frac{1}{4}k^2+\frac{9}{64}k^4+\frac{25}{256}k^6+...\right]
\eeq
or, in terms of the complementary modulus $k'=\sqrt{1-k^2}$
\beq
K(k')=\sum_{m=0}^\infty {{-1/2 \atopwithdelims ( ) m}}^2
\left[\ln \left(\frac{4}{k'}\right) - b_m\right] k'^{2m}
\eeq
where $b_0=0$ and $b_m=b_{m-1}+\frac{2}{2m(2m-1)}$. 
These, along with other similar expansions
for elliptic integrals, are used in the following:
After re-expressing all quantities in terms of $h$, 
we use the above-mentioned
series expansions for elliptic integrals to obtain an 
expression for the bending
angle (\ref{eq-exact-bangle}) in powers of $h$.  
Expanding around $h=0$ (region at infinity)
yields 
\beqa 
\alphahat (h) =4h &+& \left(-4+\frac{15}{4}\pi\right)h^2 
+\left(\frac{122}{3}-\frac{15}{2}\pi\right)h^3 
+ \left(-130+\frac{3465}{64}\pi\right)h^4 \nonumber \\ 
&+&\left(\frac{7783}{10} - \frac{3465}{16}\pi\right)h^5 + \left(-\frac{21397}{6} 
+ \frac{310695}{256}\pi\right)h^6 + \order{\left(h\right)^{7}}{}. 
\label{eq-bangle-weak} 
\eeqa 
{\it This is in exact agreement with the weak-deflection bending angle 
series found in} \cite{kp1}. 

To obtain the strong-deflection  bending angle, we expand $\alphahat(h)$ out 
from the photon sphere $h=1/3$.  It is then convenient to work in 
terms of $h'=1-3h$. The coefficient $\sqrt{r_0/Q}$ of the bending 
angle (\ref{eq-bangle-kp}) is given in terms of $h'$ by 
$$ \sqrt{\frac{r_0}{Q}} = \left[\frac{3}{3 + 4 (1 - h') h'} \right]^{1/4} $$ 
while the modulus and amplitude are 
$$ k^2 = \frac{\sqrt{3} -  2\sqrt{3} h' + \sqrt{3 
+ 4 (1 - h') h'}}{2 \sqrt{3 + 4 (1 - h') h'}} $$ 
and $$ \Psi = \sin^{-1} \sqrt{\frac{3 +  6 h' 
- \sqrt{3} \sqrt{3 + 4 (1 - h') h'}}{12 h'} }. $$ 
Since the quantity $$ k'^2 \equiv 1 - k^2 = \frac{-\sqrt{3} +  2\sqrt{3} h' 
+ \sqrt{3 + 4 (1 - h') h'}}{2 \sqrt{3 + 4 (1 - h') h'}} $$ 
approaches zero  as $h'\rightarrow 0$, it is convenient to 
re-write (\ref{eq-exact-bangle}) in terms of $k'$: 
\beq 
\label{eq-bangle-kp} 
\alphahat=4\sqrt{\frac{r_0}{Q}} \left[\, K(\sqrt{1-k'^2}) \ 
- \ F(\Psi,\sqrt{1-k'^2})\, \right]\ -\ \pi. 
\eeq 
An $h'$-series expansion for the elliptic integrals in (\ref{eq-bangle-kp}) can 
then be obtained by first expanding in terms of $k'^2$ and then expanding in $h'$. 

\medskip 
\noindent {\it Remark:} It is important to point out  that if you 
want Mathematica to compute the functions $K(\sqrt{1-k'^2})$ and 
$F(\Psi, \sqrt{1-k'^2})$ as given in our notation, then the 
commands are ${\rm EllipticK}[1- k'^2]$ and ${\rm EllipticF}[\Psi, 1- k'^2]$, 
respectively.  It is {\it incorrect} to use ${\rm EllipticK}[\sqrt{1-k'^2}]$ 
and ${\rm EllipticF}[\Psi, \sqrt{1-k'^2}]$ because Mathematica 
defines $ {\rm EllipticF}[\Psi, {\sf m}] = \int_0^\Psi \, 
(1 - {\sf m}\, \sin^2 \theta)^{-1/2}\, d \theta $ and  
$ {\rm EllipticK}[{\sf m}] = {\rm EllipticF}[\pi/2, {\sf m}]. $ 


The bending angle series for (\ref{eq-bangle-kp}) in terms of $h'$ is then 

\beqa 
\label{eq-bangle-series-hp-1} 
\alphahat + \pi &= & \Bigl[ 8\, \tanh^{-1} (\sqrt{2} - \sqrt{3}) 
+ \log(144) - 2 \log(h')\Bigr] \ + \ \frac{2 h'}{3} \nonumber \\ 
&& + \frac{1}{18} \Bigl[ -35 + 6\,\sqrt{3} + 60 \, \tanh^{-1} (\sqrt{2} 
- \sqrt{3}) + 15 \log(12) - 15 \log(h')\Bigr] \, h'^2 \nonumber \\ 
&& + \frac{1}{162} \Bigl[ 295 - 36\,\sqrt{3} - 360 \, \tanh^{-1} (\sqrt{2} 
- \sqrt{3}) - 90 \log(12) + 90 \log(h')\Bigr] \, h'^3 \nonumber \\ 
&& + \frac{1}{5184} \Bigl[ -19637 + 3420\,\sqrt{3} + 27720 \, \tanh^{-1} (\sqrt{2} 
- \sqrt{3}) + 6930 \log(12) - 6930 \log(h')\Bigr] \, h'^4 \nonumber \\ 
&& + \frac{1}{38880} \Bigl[ 211679 - 34200\,\sqrt{3} - 277200 \, 
\tanh^{-1} (\sqrt{2} - \sqrt{3}) -  69300\, \log(12) \Bigr. \nonumber \\ 
&& \hspace{4in} \Bigl.  - \ 69300\, \log(h')\Bigr] \, h'^5 \nonumber \\ 
&&  + \  \frac{1}{559872} \Bigl[ -5580415 +  989964\,\sqrt{3} 
+ 7456680 \, \tanh^{-1} (\sqrt{2} - \sqrt{3}) \Bigr. \nonumber \\ 
&& \hspace{2in} \Bigl. + \  1864170\, \log(12) - 1864170\, \log(h')\Bigr] \, h'^6 \  
+ \  \order{(h')^{7}}{}.\nonumber \\ 
\eeqa 
The series (\ref{eq-bangle-series-hp-1}) can be simplified 
significantly using the following:
 
\beqan 
8\, \tanh^{-1} (\sqrt{2} - \sqrt{3}) + \log(144) - 2 \log(h') 
&=& 2 \log\left(\frac{12(2 - \sqrt{3})}{h'} \right) \\ 
60 \, \tanh^{-1} (\sqrt{2} - \sqrt{3}) + 15 \log(12) - 15 \log(h') 
&=& 15 \log\left(\frac{12(2 - \sqrt{3})}{h'} \right) \\ 
360 \, \tanh^{-1} (\sqrt{2} - \sqrt{3}) + 90 \log(12) - 90 \log(h') 
&=& 90 \, \log\left(\frac{12(2 - \sqrt{3})}{h'}\right) \\ 
27720 \, \tanh^{-1} (\sqrt{2} - \sqrt{3}) + 6930 \log(12) - 6930 \log(h') 
&=& 6930\, \, \log\left(\frac{12(2 - \sqrt{3})}{h'}\right), 
\eeqan
 
where we made use of $ \tanh^{-1} (\sqrt{2} - \sqrt{3}) 
=  \frac{1}{4} \, \log(2 - \sqrt{3}). $ The bending angle series 
can now be expressed more simply as follows: 
\beqa 
\label{eq-bangle-series-hp} 
\alphahat &= & - \pi \ + \ 2\log{\left(\frac{12(2-\sqrt{3})}{h'}\right)} 
+ \frac{2 h'}{3} + \frac{1}{18}\left[6\sqrt{3}-35 
+15\log{\left( \frac{12(2-\sqrt{3})}{h'}\right)}\right](h')^2 \nonumber \\ 
&+&\frac{1}{162}\left[295-36\sqrt{3} 
- 90\log{\left(\frac{12(2-\sqrt{3})}{h'}\right)}\right](h')^3 \nonumber \\ 
&+&\frac{1}{5184}\left[-19637+3420\sqrt{3} 
+ 6930\log{\left(\frac{12(2-\sqrt{3})}{h'}\right)}\right](h')^4 
+ \ \order{(h')^{5}}{}. 
\eeqa 
We can extend this series arbitrarily far, but the expressions become 
cumbersome and there is no need to do so for our later invariant 
analysis. Observe that $\alphahat$ becomes arbitrarily large 
as $h'\rightarrow 0$ due to the first logarithmic term (and 
since the other logarithms are dominated by the given powers 
of $h'$). This is the reason for the terminology {\it strong deflection} 
for light rays passing near the photon sphere. 

\subsection{Comparison with Darwin's Logarithmic Bending Angle} 
\label{sec:darwin-comparison} 

The lowest order $h'$-term in (\ref{eq-bangle-series-hp}) is given by 
\beq 
\label{eq-zeroth} \frac{\alphahat + \pi}{2} \ 
= \ \log\left(\frac{12(2-\sqrt{3})}{h'}\right) \ 
+ \ \order{h'}{} \ = \  \log\left({\frac{12(2-\sqrt{3})r_0}{r_0-3 \gravr}}\right) \ 
+ \  \order{h'}{}. 
\eeq 
At first glance the reader may think that this term is the well-known 
logarithmic term found by Darwin 1959 \cite{darwin} and employed 
commonly in the literature (e.g., Eqs. (262) and (268) 
in \cite[p. 132]{chandra}). However, Darwin's result is actually 
\beq 
\label{eq-darwinterm} 
\frac{\alphahat_{\rm Darwin} + \pi}{2} 
= \log\left({\frac{36(2-\sqrt{3})\gravr}{r_0-3 \gravr}}\right). 
\eeq 
Equations (\ref{eq-zeroth}) and (\ref{eq-darwinterm}) 
are {\it not} identical!  If the quantity $r_0$ in the 
numerator of equation (\ref{eq-zeroth}) were replaced 
by $3\gravr$, we obtain Darwin's result, but this is not a 
legitimate substitution since the denominator would become 
zero.  However, in the limit where $r_0$ approaches the photon 
sphere's radius $3\gravr$, equations (\ref{eq-zeroth}) 
and (\ref{eq-darwinterm}) would both become infinitely 
large and so will be close in value in that limit. 

In Section~\ref{sec:comparison}, we
shall show explicitly how (\ref{eq-zeroth}) is a better
approximation than (\ref{eq-darwinterm}).
For the convenience of the reader, we 
review in detail in the Appendix how (\ref{eq-darwinterm}) is 
derived. Note that in the derivation of (\ref{eq-zeroth}), 
we were careful throughout to compare 
only terms at the same order, which allowed us to read off
the lowest order term directly from the series
expansion.

\subsection{Invariant Bending Angle to Third Order Beyond Logarithmic Term} 
\label{subsec-inv-3rd} 

The strong-deflection bending angle series (\ref{eq-bangle-series-hp}) 
is coordinate dependent since it is given in terms of the 
distance $r_0$ of closest approach. We now determine the 
bending angle in terms of the invariant quantity 
$$b' = 1 - \frac{\bcrit}{b},$$ 
where $\bcrit$ (critical impact parameter) is given 
by $\bcrit = 3 \sqrt{3} \gravr$. The quantity ranges 
over $[0,1]$ since $b$ increases outwards from the photon 
sphere at $b = \bcrit$ to infinity. 

In the Appendix, we derive the Darwin term (\ref{eq-darwinterm}) and
show that it is equivalent to the following --- see
equation (\ref{app-bangle-darwin-bp}): 
\beq
\label{eq-darwin-bp}
\alphahat_{\rm Darwin}  + \pi
=
\log\left[{\frac{216(7-4\sqrt{3})\, (1-b')}{b'}}\right].
\eeq

To express out bending angle in terms of the invariant $b'$, 
we first write $h'$ in terms of $b'$ using the 
relationship (\ref{r0solveb}) to get 
\beq 
\label{eq-hptob} 
h' = 1 - \frac{(1-b')}{2} \, \sec\left(\frac{1}{3}\,\cos^{-1} [-(1-b')] \right). 
\eeq 
Series expanding (\ref{eq-hptob}) in terms of $b'$ gives 
\beqa 
\label{eq-hptobseries} 
h' & = & \sqrt{\frac{2}{3}} \, (b')^{1/2} \ + \ \frac{2}{9} \, b' \ 
- \ \frac{7}{54 \sqrt{6}}\, (b')^{3/2} \ 
+ \ \frac{5}{243} \, (b')^2 \ 
- \ \frac{91}{3888 \sqrt{6}}\, (b')^{5/2} \nonumber \\ 
& & \nonumber \\
& &  \ + \ \frac{32}{6561} \, (b')^3 \ 
- \ \frac{2717}{419904 \sqrt{6}}\, (b')^{7/2} \ 
+ \ \frac{88}{59049} \, (b')^4 
\ + \ \order{(h')^{9/2}}{}. 
\eeqa

For the weak deflection bending angle series (\ref{eq-bangle-weak}), 
note that it is in a non-invariant form. To obtain an invariant 
expression, convert to the parameter $b'$ by 
using $h = (1 - h')/3$ and then employing the series 
for $h'$ in (\ref{eq-hptobseries}).  This yields 
\beqa
\label{eq-bangle-weak-bprime} 
\alphahat (b') &=& \frac{4}{3\sqrt{3}}(1-b')+\frac{5 \pi}{36}{(1-b')}^2
+\frac{128}{243\sqrt{3}}{(1-b')}^3\nonumber \\
& & \nonumber \\
& &  \ + \ \frac{385\pi}{5184}{(1-b')}^4
+\frac{3584}{10935\sqrt{3}}{(1-b')}^5
\ + \ \order{(1-b')^{6}}{}. 
\eeqa 
The first term is the Einstein term (\ref{eq-GRweak-1}) 
given in terms of $b'$ by 
\beq 
\label{eq-bangle-weak-bp-1} 
\alphahat_{\rm Einstein}(b') = \frac{4(1-b')}{3\sqrt{3}}. 
\eeq 

Turning to our strong-deflection bending angle expansion,  insert the 
series (\ref{eq-hptobseries}) into the  
$h'$-series (\ref{eq-bangle-series-hp}) for strong deflection.  This 
will yield a series in terms of the invariant $b'$: 
\beqa 
\label{eq-alpha-strong-bprime} \alphahat (b') & = & -\pi \ 
+ \ \log\Bigl[\frac{216\, (7 - 4 \sqrt{3})}{b'} \Bigr] \ 
+ \ \frac{-17 + 4\,\sqrt{3} + 5\, \log\Bigl[\frac{216\, 
(7 - 4 \sqrt{3})}{b'} \Bigr] }{18} \ b' \nonumber\\ 
& & \ + \ \frac{-879 + 236\,\sqrt{3} 
+ 205\, \log\Bigl[\frac{216\, (7 
- 4 \sqrt{3})}{b'} \Bigr] }{1296} \ (b')^2 \nonumber\\ 
& & \frac{-321590 + 90588\,\sqrt{3} + 68145\, \log\Bigl[\frac{216\, 
(7 - 4 \sqrt{3})}{b'} \Bigr] }{629856} \ (b')^3 \ \  + \ \ \order{(b')^{4}}{}. 
\eeqa 
The lowest order term
\beq
\label{eq-zeroth-bp}
\alphahat_{\rm lowest\  order}  + \pi
\equiv 
\log\left[\frac{216\, (7 - 4 \sqrt{3})}{b'} \right]
\eeq
is similar, but not identical, to the Darwin logarithmic term
(\ref{eq-darwin-bp}) 
since (\ref{eq-zeroth-bp}) takes into account 
the improvement to Darwin's result. The other terms in the 
series have not appeared in the literature before.  Although 
the series expansion of $h'$ in (\ref{eq-hptobseries}) involves 
fractional powers of $b'$, there are no fractional power terms 
in the bending angle. 

As promised in Section~\ref{subsec:affine}, we 
rewrite (\ref{eq-alpha-strong-bprime}) as an affine 
perturbation series about the logarithm: 
\beqa 
\alphahat (b') & = & \Bigl(\sigma_0 + \sigma_1 b' 
+\sigma_2 (b')^2+ \sigma_3 (b')^3 + \cdots\, \Bigr)\, 
\log \left(\frac{\lambda_0}{b'}\right) \nonumber \\ 
& & \hspace{1in}  +\Bigl(\rho_0+\rho_1 b' +\rho_2 (b')^2
+\rho_3 (b')^3 + \cdots\, \Bigr), 
\label{eq-bangle-strong-bprime} 
\eeqa 
where $\lambda_0=216(7-4\sqrt{3})$ and the $\sigma$'s 
and $\rho$'s are constants given by 
\beqan 
\sigma_0&=&1\;\;\;\;\;\;\;\;\;\;\;\;\;
\;\;\;\;\;\;\;\;\;\;\; {\rho_0}=-\pi\\ 
\sigma_1&=&\frac{5}{18}\;\;\;\;\;\;\;\;
\;\;\;\;\;\;\;\;\;\;\;\;\; {\rho_1}=\frac{-17+4\sqrt{3}}{18}\\ 
\sigma_2&=&\frac{205}{1296}\;\;\;\;\;\;\;\;
\;\;\;\;\;\;\;\;\;\; {\rho_2}=\frac{-879+236\sqrt{3}}{1296}\\ 
\sigma_3&=&\frac{68145}{629856}\;\;\;\;\;\;
\;\;\;\;\;\;\;\;\; {\rho_3}=\frac{-321590+90588\sqrt{3}}{629856}. 
\eeqan 
The terminologies below will be employed for the terms of the 
affine series (\ref{eq-bangle-strong-bprime}): 
\beqan 
&& \mbox{$0$th-order:}\quad \sigma_0 \, \log \left(\lambda_0/b'\right) 
+ \rho_0 \\ && \mbox{$1$st-order:}\quad \left(\sigma_0 
+ \sigma_1 b' \, \right)\, \log \left(\lambda_0/b'\right) \ 
+ \  \left(\rho_0 + \rho_1 b' \right) \\ 
&& \mbox{$2$nd-order:}\quad \left(\sigma_0 + \sigma_1 b' 
+\sigma_2 (b')^2 \, \right)\, \log \left(\lambda_0/b'\right) \ 
+ \  \left(\rho_0 + \rho_1 b'  +\rho_2 (b')^2 \,\right) \\ 
&& \mbox{$3$rd-order:}\quad \left(\sigma_0 + \sigma_1 b' 
+\sigma_2 (b')^2+ \sigma_3 (b')^3 \right)\, \log \left(\lambda_0/b'\right) \ 
+ \ \left(\rho_0+\rho_1 b' +\rho_2 (b')^2+\rho_3 (b')^3 \right). 
\eeqan 
We have chosen to truncate at $3$rd-order because this order 
will already probe accurately as far as to twice the critical 
impact parameter.  These issues are taken up in the next section. 

\section{Comparison of Perturbative and Exact Bending Angles} 
\label{sec:comparison} 

This section will give a numerical comparison of the formal 
exact bending angle with the zeroth, $1$st-, $2$nd- and $3$rd-order 
affine corrections to the logarithmic term. The comparison will be 
given by the following percentage discrepancy: 
$\displaystyle \left(\frac{\alphahat_{\rm exact}-\alphahat}
{\alphahat_{\rm exact}}\right) \times 100\, \%.$ 

\begin{figure}[htp] 
\begin{center} 
\epsfxsize=4in \epsfysize=2.5in {\epsfbox{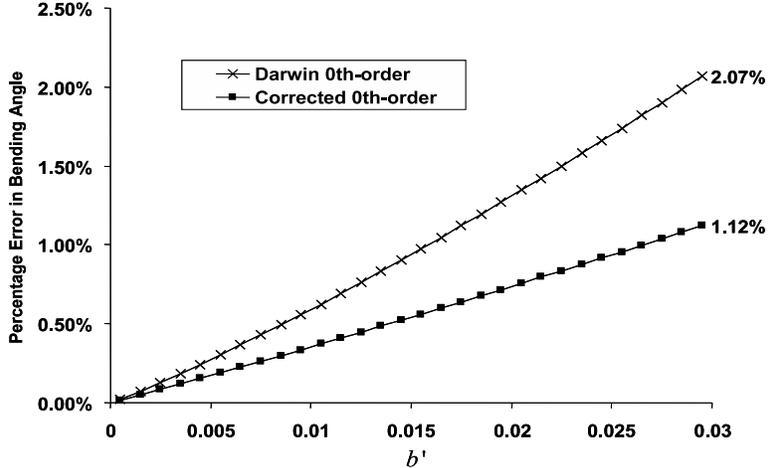}} 
\caption{\small \sl Percentage Discrepancy between Darwin's 
logarithmic term, our $0$th-order logarithmic term, and the 
exact numerical value represented by the horizonatal 
axis. Our $0$th-order term is a better approximation to the exact value.} 
\label{fig:PercentDarwin0} 
\end{center} 
\end{figure} 

In Figure~\ref{fig:PercentDarwin0} is plotted the percentage 
discrepancy between the Darwin term (\ref{eq-darwin-bp}) 
written in terms of $b'$, our $0$th-order term (\ref{eq-zeroth}), 
and the exact numerical value (\ref{eq-exact-bangle}), which is 
represented by the horizontal axis. The $0$th-order term is 
closer to the exact value and comes within $1\%$ of the 
exact value at  $b' \approx 0.03$. The deviations increase 
for larger values of $b'$. 

\begin{figure}[htp] 
\begin{center}  
\epsfxsize=4.5in \epsfysize=2.75in{\epsfbox{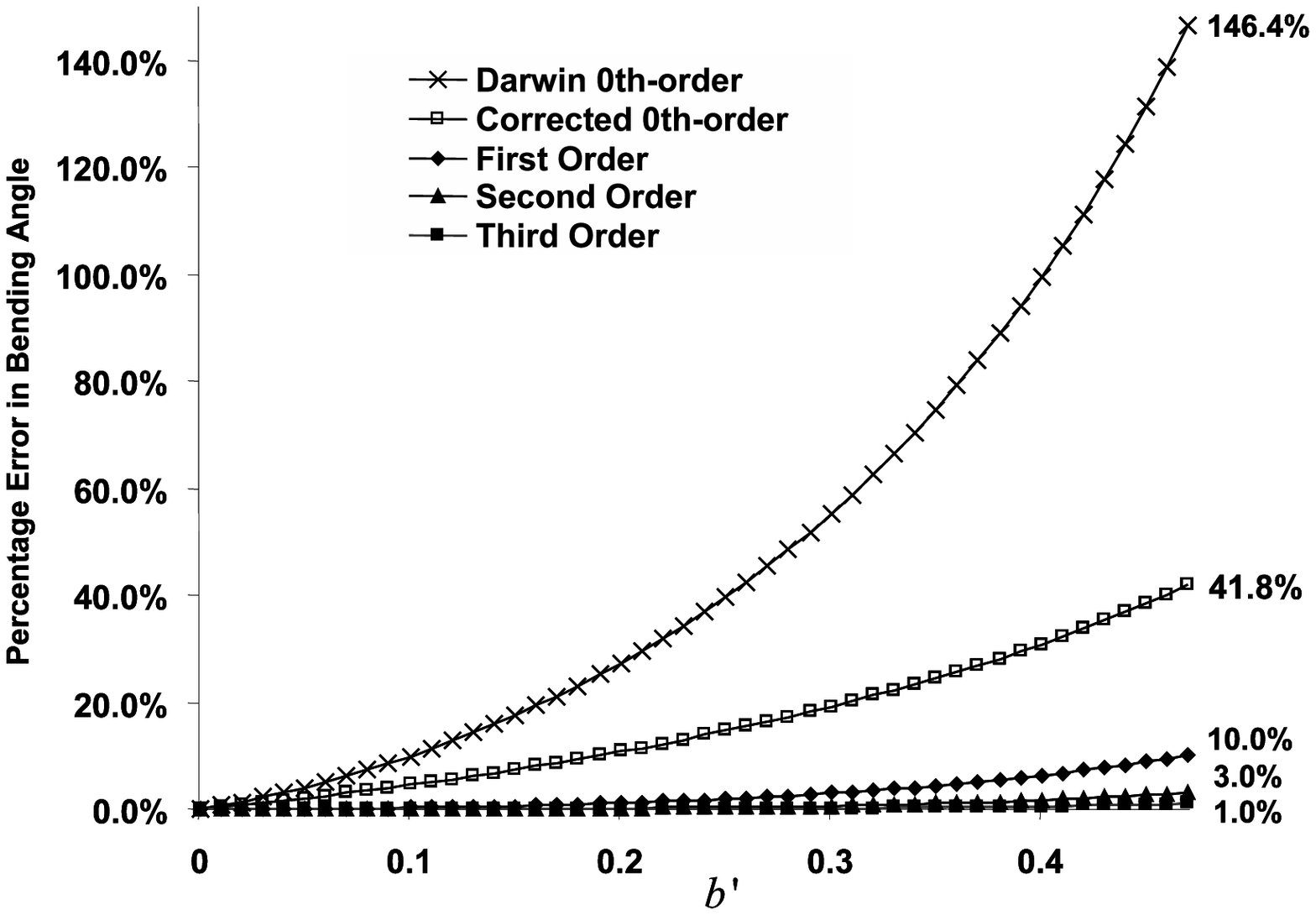}}
\epsfxsize=4.5in \epsfysize=2.75in{\epsfbox{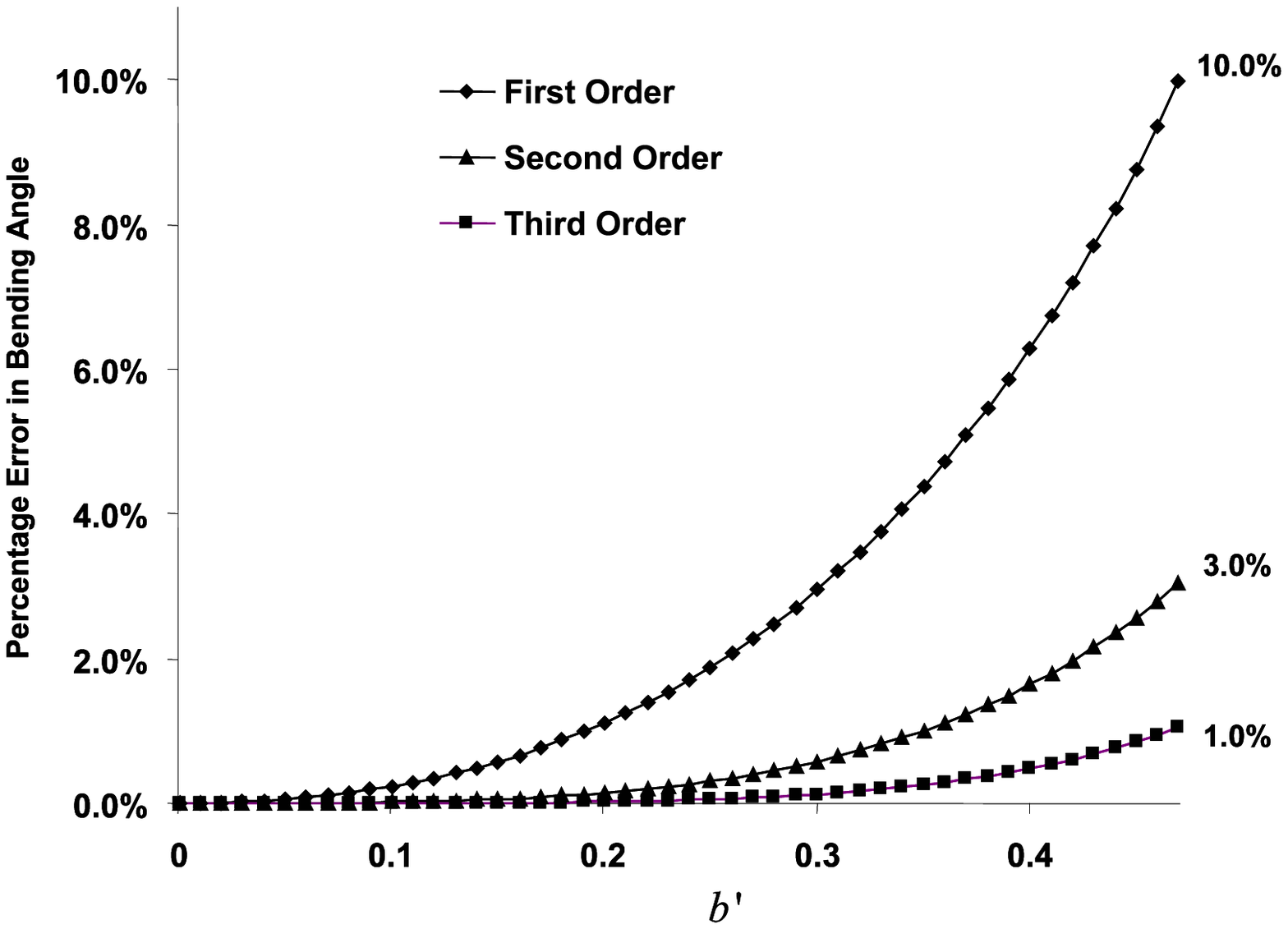}} 
\caption{\small \sl Top graph: Percentage discrepancy between 
the $0$th- to $3$rd-order terms, the Darwin logarithmic term, and 
the exact result given by the horizontal axis.  The $3$rd-order 
result deviates by at most $1\%$ from the exact value with the 
maximum deviation at $b'\approx 0.4705$. Bottom graph: A close-up of the 
same graph is plotted to show clearly the higher order corrections in more detail.}  
\label{fig:PercentDarwin0123}  
\end{center} 
\end{figure}

If we include terms through to $3$rd-order in the strong 
deflection expression (\ref{eq-bangle-strong-bprime}), then 
unlike the case in Figure~\ref{fig:PercentDarwin0}, the 
series comes to within $1\%$ of the exact value for 
regions well past $b' \approx 0.03$. Figure~\ref{fig:PercentDarwin0123} shows 
that the regions could be as far out as to roughly {\it twice the critical 
impact parameter} (i.e., $b \approx 2 \bcrit$ or $b' \approx 0.4705$) and 
still be within a $1\%$ discrepancy. Beyond this point the discrepancy 
continues to increase and will not be shown.
 
\begin{figure}[htp] 
\begin{center}  
\epsfxsize=4.5in \epsfysize=2.75in {\epsfbox{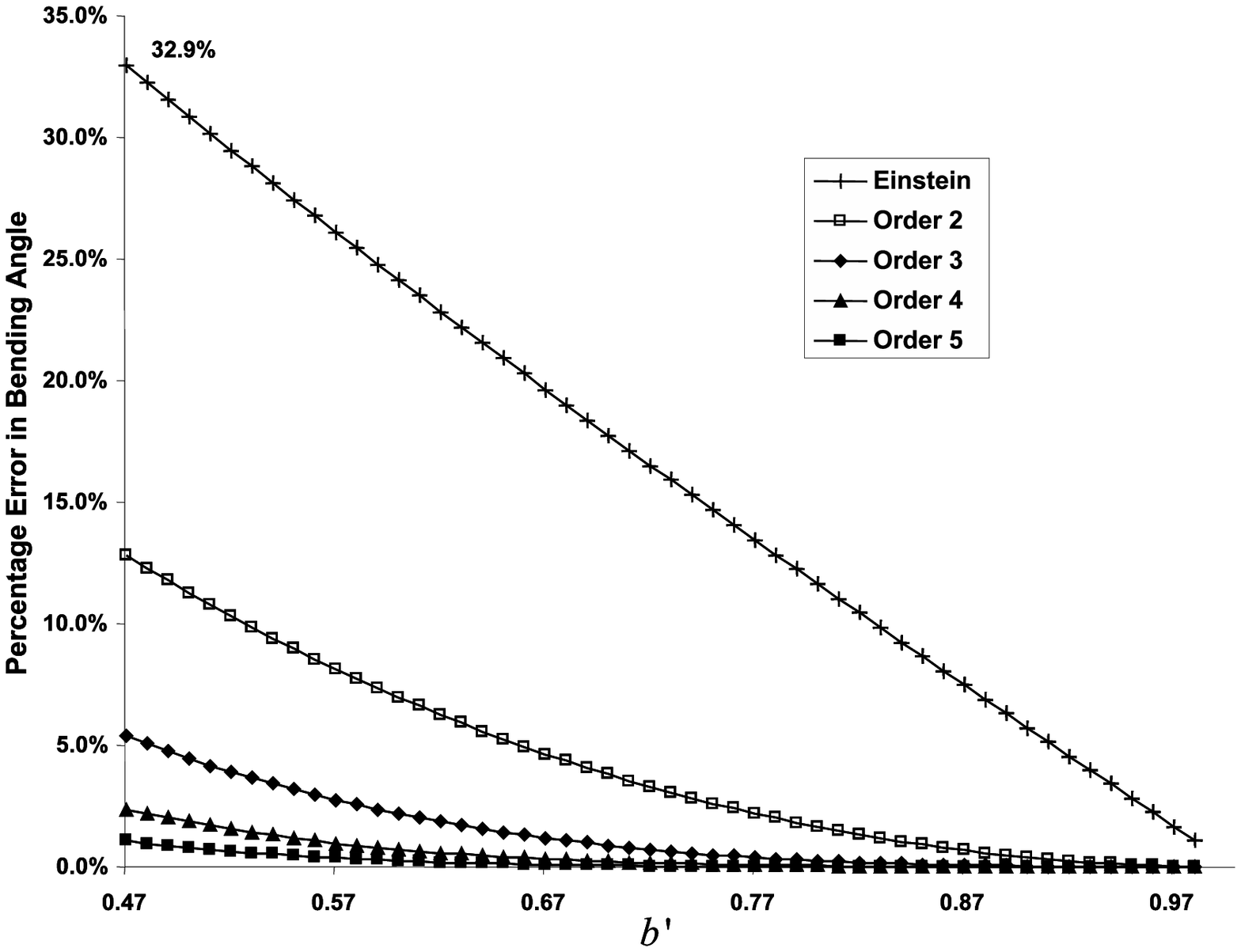}}  
\caption{\small \sl Percentage discrepancy corresponding to 
various orders of the weak deflection bending angle. The 
horizontal axis corresponds to the exact value. 
The $5$th-order expansion has a $1\%$ discrepancy at $b' 
\approx 0.04705$.} 
\label{fig:PercentOutEinsteinAll}  
\end{center} 
\end{figure}

For the $5$th-order weak deflection bending angle 
series (\ref{eq-bangle-weak-bprime}) and the Einstein 
term (\ref{eq-bangle-weak-bp-1}), Figure~\ref{fig:PercentOutEinsteinAll} 
shows the percentage discrepancies until (\ref{eq-bangle-weak-bprime}) 
reaches a $1\%$ deviation.  This occurs at $b' \approx 0.4705$ and has 
a larger deviation for greater $b'$ values. 

\begin{figure}[htp] 
\begin{center}  
\epsfxsize=4.5in \epsfysize=2.75in {\epsfbox{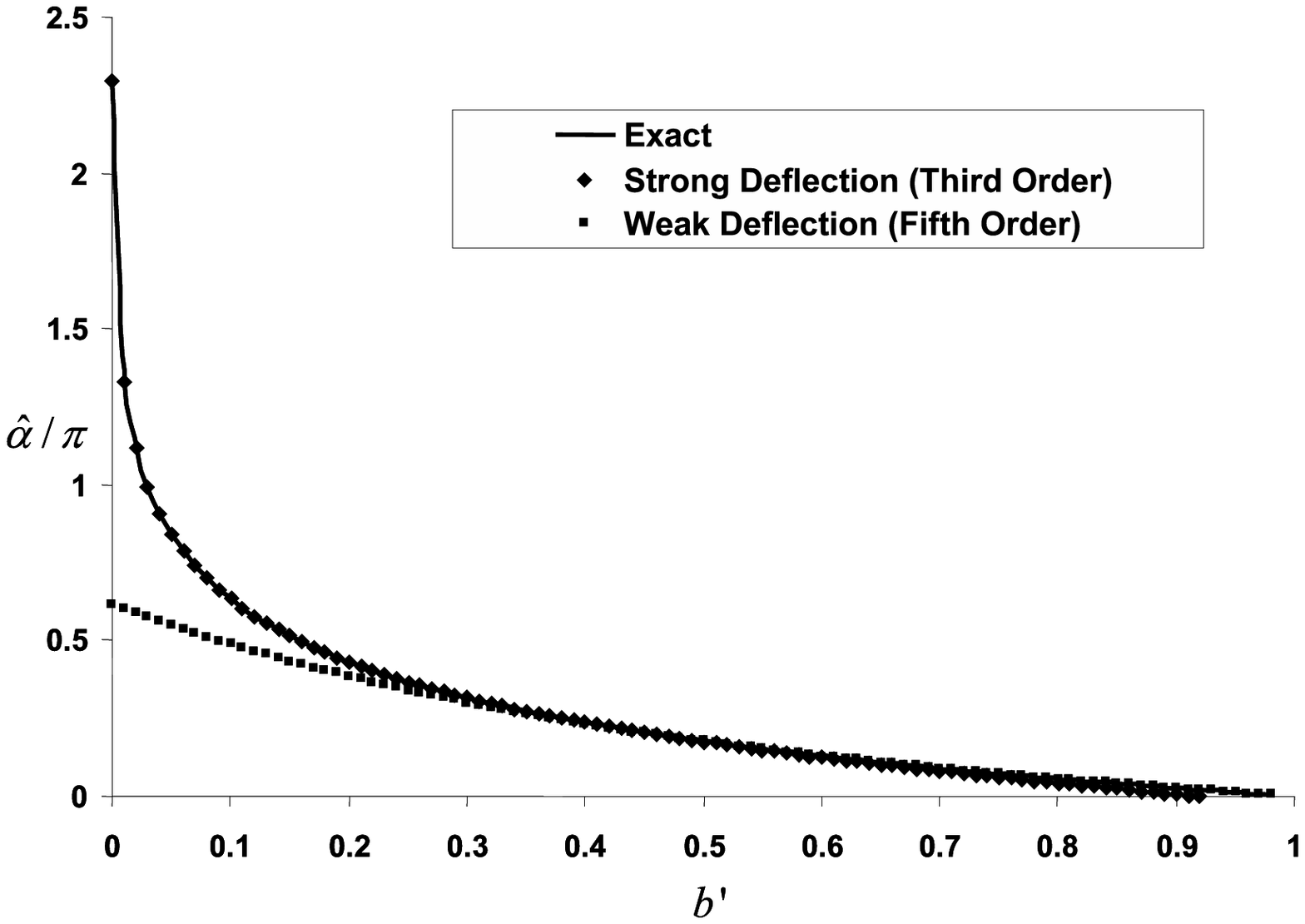}}  
\caption{\small \sl The 3rd-order strong deflection and the 
5th-order weak deflection are plotted alongside the numerically
integrated exact formal bending angle.}  
\label{fig:AlphaIn3Out5}  
\end{center} 
\end{figure}

\begin{figure}[htp] 
\begin{center}  
\epsfxsize=4.5in \epsfysize=2.75in {\epsfbox{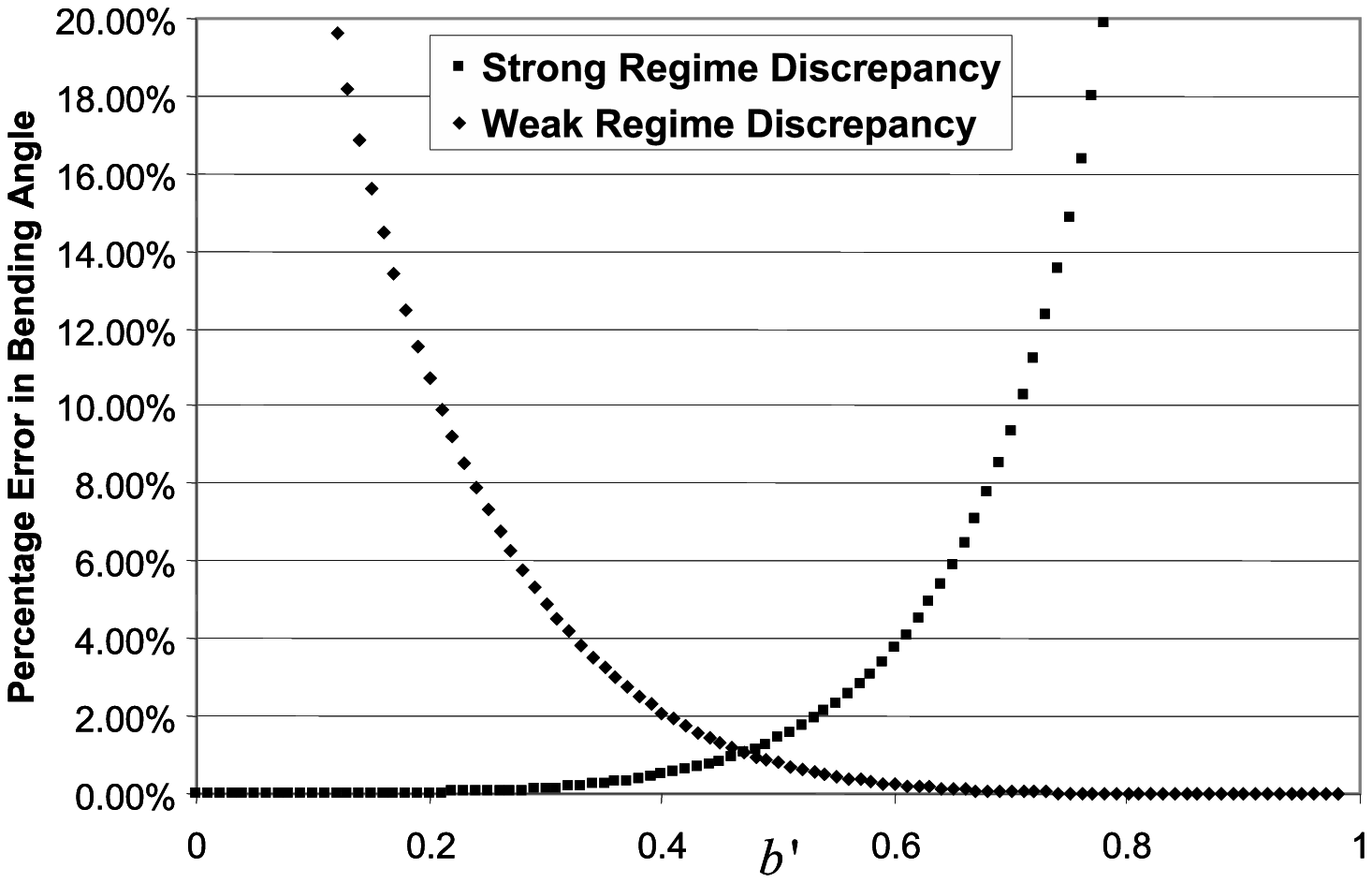}}  
\caption{\small \sl Percentage discrepancy for $3$rd-order 
strong deflection and $5$th-order weak deflection. 
The horizontal axis corresponds to the exact value.  The 
two discrepancies criss-cross essentially at $1\%$ 
for $b' \approx 0.4705$.}  
\label{fig:SWUnderOne}  
\end{center} 
\end{figure}

Figure~\ref{fig:AlphaIn3Out5} shows both the $3$rd-order 
strong deflection (\ref{eq-bangle-strong-bprime}) and $5$th-order 
weak deflection (\ref{eq-bangle-weak-bprime}) plotted alongside 
the exact result. The same comparison, in terms of the percentage 
discrepancy is presented in Figure~\ref{fig:SWUnderOne}.  This graph 
illustrates that if (\ref{eq-bangle-strong-bprime}) is used on the interval 
$0 < b' \lesssim 0.4705$ and (\ref{eq-bangle-weak-bprime}) 
on the interval $0.4705 \lesssim b' < 1$, then together these 
two series deviate by at most $1\%$ from the exact value.  In addition, 
the $1\%$ maximum deviation occurs at $b' \approx 0.4705$. These two 
series can then yield bending angle information rather accurately 
from the photon sphere to infinity. 

\section{Conclusions} 

An analytical perturbation framework 
for calculating the bending angle of light rays traversing 
the gravitational field of a Schwarzschild black hole was 
given. We expressed the strong-deflection bending angle in 
an invariant affine perturbation series in $b'$ about a 
logarithmic function. Our logarithmic expression was shown 
to be a better approximation to the exact bending  angle 
than the logarithmic one found by Darwin, which is commonly
used in the literature. We also derived 
the known weak deflection bending angle series as a 
consistency check of our framework. The latter series 
was reformulated in terms of the more natural perturbation 
parameter $b$, which smoothly transitions from strong deflection 
near the photon sphere to weak deflection at infinity. 
Comparison was then given of our invariant strong deflection 
bending-angle series  with the numerically integrated exact 
formal bending angle expression. We found less than $1 \%$ discrepancy 
for light rays as far out  as  twice the critical impact parameter. 
This was followed by a further comparison with our invariant form of 
the weak deflection series. It was found that taken together, the two series 
yield an approximation that is within $1 \%$   
of the exact bending angle value for light rays 
traversing anywhere between the photon sphere and infinity. 

\begin{acknowledgments} 

This work was supported by 
NSF grants DMS-0302812, AST-0434277, and AST-0433809.  
The authors acknowledge the 2005 AIM workshop on Kerr black holes, 
where this collaboration was started. S. I. thanks the 
Department of Mathematics at Duke University for their 
hospitality and SUNY-Geneseo for a Mid-Career Summer grant. 
\end{acknowledgments} 

\appendix 

\section{Derivation of the Darwin Term} 

For the convenience of the reader, we present here the standard
bending angle derivation as in \cite[p. 132]{chandra}. The formal 
exact bending angle $\alpha$ is given (\ref{eq-exact-bangle}) 
by 
$$ \frac{1}{2}(\pi+\alphahat) = 2\left(\frac{r_0}{Q}\right)^{1/2} 
\left[K(k)-F(\Psi,k)\right], $$ 
where 
$$ Q  =  \sqrt{(r_0-2\gravr)(r_0+6\gravr)},  
\quad k^2  =   \frac{Q - r_0 + 6\gravr}{2 Q},  
\quad \Psi  =  \sin^{-1} 
\sqrt{\frac{Q  + 2 \gravr - r_0}{Q  + 6 \gravr - r_0}}. $$ 
For the reader's convenience,  the correspondence between our 
notation and that in \cite[p. 132]{chandra} is as 
follows:  $(\pi+\alphahat)/2 = \phi_\infty$, 
$\alphahat = \Theta$, $P = r_0$, $D=b$, and $D_c = b_c$. 

The perturbation scheme in \cite[p.132]{chandra} is defined 
by setting 
\beq 
\label{eq-r0-1} r_0=\gravr(3+\delta). 
\eeq Note that $\delta$ does not remain under the 
value $1$ as one moves away from the photon sphere.  
The value of $\delta$ is assumed to be small compared 
to $3$.  This is important for the  approximations to follow. 
Series expanding and keeping terms at $1$st-order 
in $\delta$ yields 
$$ Q=\sqrt{(r_0-2\gravr)(r_0+6\gravr)} =\sqrt{9\gravr^2
+10 \delta \gravr^2+\delta^2 \gravr^2} =3\gravr \left[1+\frac{10 \delta}{9} 
+ \frac{\delta^2}{9}\right]^{1/2}. $$ 
Consequently, 
$$ Q=\gravr \left(3+\frac{5\delta}{3}\right). $$ 
Similarly, 
\beqan 
{k'}^2=1-k^2&=&1-\frac{Q-r_0+6\gravr}{2Q}\\ 
&=&1-\frac{\gravr \left(3+\frac{5\delta}{3}\right) 
-\gravr(3+\delta)+6\gravr}{2\gravr \left(3+\frac{5\delta}{3}\right)}\\ 
&=&1-\frac{1}{6}\left[6-\delta+\frac{5\delta}{3}\right] 
\left(1-\frac{5\delta}{9}\right)^{-1}, 
\eeqan 
which gives 
\beq
\label{kprime-delta}
{k'}^2=\frac{4}{9}\delta.
\eeq
Note that the quantity $\delta$ is now related to the 
complementary modulus of the elliptic function allowing for a 
series expansion in terms of $k'$. If we consider the limit as
$k' \rightarrow 0$ we get \cite[p. 132]{chandra}: 
$$ K(k) = K(\sqrt{1- {k'}^2})\rightarrow \log{\left(\frac{4}{k'}\right)} 
= \frac{1}{2}\log{\left(\frac{16}{{k'}^2}\right)} 
=  \frac{1}{2}\log{\left(\frac{16}{4\delta /9}\right)} 
= \log{6}-\frac{1}{2}\log{(\delta)} $$ 
and 
$$ F(\Psi,k) 
= F(\Psi, \sqrt{1- {k'}^2})\rightarrow 
\frac{1}{2}\log{\left(\frac{\sqrt{3} + 1}{\sqrt{3} - 1}\right)} $$ 
since $\Psi \rightarrow \sin^{-1}(1/\sqrt{3})$. 
Taken together, these results yield 
\beq 
\label{app-bangle-darwin-1} 
\frac{1}{2}(\pi+\alphahat) \rightarrow 
\frac{1}{2}\log{\left[\frac{6^4 \, \sqrt{3}\, 
(\sqrt{3} - 1)^2}{ 2 (\sqrt{3} + 1)^2}\right]} 
- \frac{1}{2}\log{\left[\frac{\sqrt{3}}{2}\, \gravr \, \delta^2\right]}. 
\eeq 
In the next step, the $\delta$ is solved for in terms of $r_0$ and $\gravr$
using the first-order scheme (\ref{eq-r0-1}), 
and then substituted in the second-order $\delta^2$
in (\ref{app-bangle-darwin-1}) to get the leading 
term in equation (268) of \cite[p. 132]{chandra}:
\beqa 
\frac{1}{2}(\pi+\alphahat) \rightarrow && \frac{1}{2}
\log{\left[\frac{6^4 \, \sqrt{3}\, (\sqrt{3} - 1)^2}{ 2 (\sqrt{3} + 1)^2}\right]} 
- \frac{1}{2}\log{\left[\frac{\sqrt{3}}{2}\, 
\frac{(r_0 - 3 \gravr)^2}{\gravr^2}\right]}  \nonumber \\ 
& & = 
\log \left[\frac{36 (2 - \sqrt{3})\, \gravr}{r_0 \, - \, 3 \gravr}\right]. 
\label{app-bangle-darwin-3} 
\eeqa 
It would seem that the first-order scheme (\ref{eq-r0-1}) should have
been extended to $\delta^2$ originally if second-order terms in $\delta$
would be considered in the analysis.  
However, {\it we leave it to the judgment of the reader to decide whether
the mixing of first and second order terms is appropriate in the
above derivation.}
Finally, note that equation (\ref{app-bangle-darwin-3}) is the Darwin term
quoted at the beginning of the paper --- see (\ref{eq-GRstrong-1}).

We can re-express (\ref{app-bangle-darwin-3}) 
in terms of $\gravr$ and the impact parameter
$b$. By equation (\ref{r0tob}), we have
\beq
\label{app-eq-btor0}
b = \frac{r_0^{3/2}}{\sqrt{r_0 - 3\gravr}}.
\eeq
Insert the first-order $\delta$ equation 
(\ref{eq-r0-1})  
into (\ref{app-eq-btor0}) 
and expand to second-order in the perturbation parameter
$\delta$:
\beq
\label{app-eq-bseries-delta}
b = \sqrt{
\frac{\gravr^3 \, (3 + \delta)^3}{\gravr \, (3 + \delta) - 2 \gravr}
}
= 3 \sqrt{3}\gravr \ +  \ 
\frac{\sqrt{3}}{2} \gravr \, \delta^2 \ + \ {\cal O}(\delta^3).
\eeq
Writing (\ref{app-eq-bseries-delta}) to  second-order in $\delta$
yields  the result in equation
(263) of  \cite{chandra}:
\beq
\label{app-eq-b-bc}
b - b_c
=
\frac{\sqrt{3} \gravr \delta^2}{2}.
\eeq
{\it Once again,
we leave it to the reader to decide whether
the mixing of first and second order terms is appropriate in the
above derivation of} (\ref{app-eq-b-bc}).
Now,
substituting  
$$
\delta = \frac{r_0 - 3 \gravr}{\gravr}
$$
from (\ref{eq-r0-1}) into 
(\ref{app-eq-b-bc}) yields
$$
\left(\frac{\gravr}{r_0 - 3 \gravr}\right)^2
= \frac{\sqrt{3}}{2} \frac{\gravr}{(b- b_c)}.
$$ 
Using 
$$
(2 - \sqrt{3})^2 
=
7 - 4\, \sqrt{3}
=
\frac{\left(\sqrt{3}-1\right)^2}{\left(\sqrt{3}+1\right)^2},
$$
we find from (\ref{app-bangle-darwin-3}) that
\beq
\label{app-bangle-darwin-2}
\pi+ \alphahat =
\log \left[36^2 (2 - \sqrt{3})^2\, 
\left(\frac{\gravr}{r_0 \, - \, 3 \gravr}\right)^2\right]
=
\log{\left[
\frac{648 \sqrt{3}\, \left(\sqrt{3}-1\right)^2\, \gravr}{
\left(\sqrt{3}+1\right)^2 \, (b-b_c)}\right]}.
\eeq
Equations 
(\ref{app-bangle-darwin-3}) and  (\ref{app-bangle-darwin-2}) 
are among the common forms used in the literature \cite{bangleform},
and both are based on combining first- and second-order terms
in $\delta$.
Finally, we can also express the Darwin term using the variable $b'$.
Since 
$$ 
\frac{3 \sqrt{3} \, \gravr}{b - b_c}
= \frac{1- b'}{b'}, 
$$
we see that equation (\ref{app-bangle-darwin-2}) is equivalent to
\beq
\label{app-bangle-darwin-bp}
\pi + \alphahat =
\log{\left[\frac{216 (7 - 4 \sqrt{3}) \, (1 - b')}{b'} \right]}.
\eeq

Our study in 
Section~\ref{subsec:Bangle-series-hp} carries out the perturbation 
analysis for obtaining $\alphahat$ consistently, matching 
terms of the same order.  This 
yields a different expression for the leading logarithmic term,
one that is more accurate.

\end{document}